# Imposition of Different Optimizing Object with Non-Linear Constraints on Flux Sampling and Elimination of Free Futile Pathways

Lu Xie and Yi Zhang

Life Science School, University of Science and Technology of China, Hefei, China

**Abstract**

Constraint-based modeling has been widely used on metabolic networks analysis, such as biosynthetic prediction and flux optimization. The linear constraints, like mass conservation constraint, reversibility constraint, biological capacity constraint, can be imposed on linear algorithms. However, recently a non-linear constraint based on the second thermodynamic law, known as "loop law", has emerged and challenged the existing algorithms. Proven to be unfeasible with linear solutions, this non-linear constraint has been successfully imposed on the sampling process. In this place, Monte - Carlo sampling with Metropolis criterion and Simulated Annealing has been introduced to optimize the Biomass synthesis of genome scale metabolic network of *Helicobacter pylori* (*i*IT341 GSM / GPR) under mass conservation constraint, biological capacity constraint, and thermodynamic constraints including reversibility and "loop law". The sampling method has also been employed to optimize a non-linear objective function, the Biomass synthetic rate, which is unified by the total income number of reducible electrons. To verify whether a sample contains internal loops, an automatic solution has been developed based on solving a set of inequalities. In addition, a new type of pathway has been proposed here, the Futile Pathway, which has three properties: 1) its mass flow could be self-balanced; 2) it has exchange reactions; 3) it is independent to the biomass synthesis. To eliminate the fluxes of the Futile Pathways in the sampling results, a linear programming based method has been suggested and the results have showed improved correlations among the reaction fluxes in the pathways related to Biomass synthesis.

1. INTRODUCTION

As a constraint-based model, Flux Balance Analysis (FBA) (1) has been used widely (2) to investigate the structure and functioning of metabolic networks. The flux vectors satisfying the steady-state hypothesis constitute a solution space, on which different constraints could be imposed to implement different objectives. The imposition of linear constraints, like physicochemical reversibility and biological capacity, could shape the solution space into a convex cone (3), where linear programming can be used to optimize linear objectives, such as predicting the optimal growth rates (4), measuring ranges of achievable flux values (5), and minimizing the stationary metabolic fluxes (6). However, the imposition of non-linear constraints, or the optimization of non-linear objectives, requires other optimization methods than linear programming. At this point, sampling-based methods have the advantages in the imposition of non-linear constraints or optimization objectives, and have been employed in many fields of metabolic network analysis (7). Besides, sampling-based methods are also convenient of applying post processing techniques (7).

In this work, a Monte Carlo sampling-based process has been proposed to optimize the Biomass synthesis flux and Biomass synthetic efficiency of the genome-scale metabolic network of *H. pylori* (*i*IT341 GSM/GPR) (8) through sampling in the constrained solution space. Aside from the linear constraints above, a non-linear constraint, termed as the "Loop Law" (9-10), has also been imposed on the sampling process. After the imposition, the solution space has been separated into discontinuous sub-spaces, identified by the direction patterns of the reversible reactions in the loop, and the flux vectors in each of the sub-space could satisfy the constraint of "loop law" (10-11). For the discontinuity of the solution space, sampling method has successfully been employed to analysis the metabolic network of *H. pylori* (12). As an advantage of sampling methods, the optimization of non-linear objective could be introduced as the process of searching the minima of an energy function (13-14). In our approach, Metropolis criterion has been introduced to optimize the Biomass production through the sampling process with Simulated Annealing technique (15).

As a key negative feedback of Monte - Carlo sampling, there is currently no ideal method to measure the coverage(7). For this reason, Artificial Centered Hit and Run (ACHR) algorithm (16-17) has been used here to facilitate the sampling progress with multiple random start points. In addition, the consistency of those sample trajectories with different start point could be evaluated by calculating the Cumulative Mean Square Inner Product (CMSIP) (18) between each two trajectories. CMSIP can also be used to evaluate the convergence of sampling, by comparing the consistency of each two adjacent fractions of one entire trajectory.

To compensate the missing data of biological capacities of some certain reactions, the upper and lower flux boundary of these reactions are usually fixed artificially. As a consequence of that, the maximum value of Biomass synthesis flux would probably rely on these artificial constraints. One attempt to extenuate the reliance is flux minimization (6), in which the flux value was unified by the Gibbs' Free Energy ($\Delta G$). However, collecting the $\Delta G$ data set of each substrate also stands as a challenge for a genome-scale network. In our method, the Biomass synthesis flux has been divided by the total amount of reducible electrons from all income fluxes, as a measure of the Biomass synthetic efficiency optimization.

Deeper investigation of the metabolic network has reaffirmed the necessity of flux minimization. Similar to the definition of internal loop, a new kind of pathway, the Futile Pathway, has been proposed here (Fig. 1). The Futile Pathways are self-balanced sub networks and contribute nothing to the Biomass synthesis, while the difference from internal loops is that the Futile Pathways include exchange reactions. Most Futile Pathways, whose substrates contain reducible electrons, could be constrained during the optimization of Biomass synthetic efficiency. In contrast, the other Futile Pathways, or named Free Futile Pathways (FFP), can only be constrained by their biological capacities, and the existence of fluxes in these pathways might cover the correlations among other fluxes which indicate functional pathways (19). As a post sampling processing, the fluxes in the FFP have been minimized by Linear Programming.

## 2. MATERIAL AND METHOD

### 2.1. System and Model

As a kind of notorious human gastric pathogen, *H. pylori* infection has been detected in more than half cases of gastric and duodenal ulcers (20). The genome-scale metabolic network of the *H. pylori* strain 26695 has been reconstructed using the revised genome annotation and new experimental data in 2005 and marked as *i*IT341 GSM/GPR (8), which accounts for 341 metabolic genes, 476 internal reactions, 74 external reactions, 8 demand functions, 411 internal metabolites and 74 external metabolites. The demand functions are the reactions of the type: A -->, which means that the compound A can be only produced by the network, but without further balancing the compound A. The demand functions include a Biomass function, a HMFURN function, excretion of Thiamin, Menaquinone 6, Biotin and Heme (Protoheme), sinking reactions of ahcys(c) and amob. The external reactions and demand functions, or called exchange reactions, serve as the input and output of the network. The raw version of stoichiometric matrix, in total, has 558 reactions and 485 metabolites.

### 2.2. Network Preprocessing

#### 2.2.1. Deleting the Zero Flux Reactions

The original model *i*IT341 GSM/GPR contains dead-end metabolites and zero fluxes (8). One way to find these reactions is to calculate the maximum and the minimum values of every flux by Linear Programming (5),

$$
\begin{aligned}
&for(i \in R) \\
&\quad \min(-v_i) \,\&\, \min(v_i) \\
&\quad s.t. \begin{cases} S \cdot v = 0 \\ \alpha_k \leq v_k \leq \beta_k, k \in R \end{cases} \\
&end
\end{aligned}
\qquad (1)
$$

where $S$ stands for the stoichiometric matrix, $R$ stands for the set of reactions, $\alpha_k$ and $\beta_k$ denote the lower and upper bound of the $k$-th flux, respectively. The $v_i$ which is always tightly constrained to zero reflects that the i-th reaction contains at least one dead-end metabolite. Deletion of these reactions and the dead-end metabolites would reduce the dimension of the stoichiometric matrix and the size of solution space.

*2.2.2. Locating the Internal Loops*

Prior to avoiding the existence of fluxes in the internal loops, the combinations of those reactions that are capable of forming loops must be located first. As described above, the internal loops do not contain exchange fluxes (11). For this character, the set of exchange reactions, noted as $R_{ex}$, were forced to zero in order to make the system closed. After that a Linear Programming was performed to find the maximum and minimum allowable fluxes on the remained reactions, each at a time (5).

$$for(i \in \overline{R^{ex}})$$
$$\min(-v_i) \ \& \ \min(v_i)$$
$$s.t. \begin{cases} S \cdot v = 0 \\ v_j = 0, j \in R^{ex} \\ \alpha_k \leq v_k \leq \beta_k, k \in \overline{R^{ex}} \end{cases} \quad (2)$$
$$end$$

Here the bar over a set means its complementary set. From the results, the fluxes of the reactions that are unable to form loops were constrained to zero while others were not, and these non-zero reactions may form one or several loops. To determine the least reactions for each loop, another Linear Programming is needed.

$$for(i \in R^{loop})$$
$$\min(\sum_{j \neq i | j \in R} |v'_j|)$$
$$s.t. \begin{cases} v'_i = v_i \\ S \cdot v' = 0 \end{cases} \quad (3)$$
$$end$$

For each loop flux $v_i$, the fluxes that remained non-zero in $v'$ are necessary for $v_i$ to form its belonging loop(s). Those reactions, each of which merely belongs to one loop, would help us to identify each of these loops.

*2.2.3. Finding Feasible Direction Patterns*

A loop is an set of reactions, noted as $R^{loop}$, which compose a sub-network that could be mass-balanced but contains no exchange reactions with the environment. The sum

of the changes in chemical potentials around a loop equal zero:

$$\sum \Delta u_i = 0, i \in R^{loop} \tag{4}$$

where

$$\Delta u_i = \sum S_{j,i} \cdot u_j, i \in R, j \in M \tag{5}$$

Here $\Delta u$ is a vector whose element indicates the chemical potential change of each reaction, and u is a vector whose element indicates the chemical potential of each metabolite. S is the stoichiometric matrix, whose rows and columns represent the metabolites and reactions, respectively. For each reaction, the direction of its flux and the change of chemical potential should satisfy the constraint based on the laws of thermodynamics, i.e., all fluxes must follow the downward direction of the chemical potential change:

$$v_i \cdot \Delta u_i \leq 0, \forall i \in R \tag{6}$$

Here $v$ denotes for the flux vector and $v_i$ denotes for the $i^{th}$ flux value. When the inequalities Eq. 6 are satisfied, the net flux around a biochemical loop must be forced to zero, which reflects the essence of "loop law".

As cited before, the formation of a loop relies on the direction patterns of the reversible reactions belonging to this loop (10-11). For a specific loop, noted as $R_s^{loop}$, which contains r reversible reactions, has $2^r$ possible direction patterns. To examine which patterns cannot lead into loop, a number of $2^r$ flux vectors were generated, each corresponding to a unique direction pattern. Every flux vector satisfies the stoichiometric constraint and the biological capacity constraint, and the flux values were randomly generated. These vectors, noted as ($v^t \mid t = 1\ldots 2^r$), were lately examined by solving a set of inequalities each at a time.

$$\begin{aligned} &for(t = 1\ldots 2^r) \\ &\quad solve[u \mid \sum_{j \in M} u_j \cdot (S_{j,i} \cdot v_i^t) < 0; i \in R^{rev} \cap R_s^{loop}] \\ &end \end{aligned} \tag{7}$$

Here $R^{rev}$ denotes the set of reversible reactions and the vector u represents the set of virtual chemical potential of the metabolites. If the set of inequality is solvable under a specific direction pattern, there must be a set of feasible chemical potentials for the metabolites to satisfy the "Loop Law", which means this pattern cannot form loop.

**2.3. Monte-Carlo sampling**

*2.3.1. The Sampling Space and Method*

Given the stoichiometric matrix *S*, with *m* rows and *n* columns, if the rank of *S* is smaller than *n*, there is a null space (16, 21) in which all the flux vectors satisfy the linear equation Eq. 3. The base vector set of the null space, noted as *N*, has n rows and *n - rank (S)* columns and each of its columns corresponds to a base vector of the null space. The sampling was performed in the null space and its objectives are the coordinates of the flux vectors, noted as *c*.

$$v = N \cdot c \qquad (8)$$

To impose the biological constraint, each flux value has a lower limit $\alpha$ and upper limit $\beta$. For the reactions with known biological capacity, the values of $\alpha$ and $\beta$ were chosen as the Ref. (12). For others, values of $\alpha$ and $\beta$ were set to +/-100, respectively, and $\alpha \geq 0$ specifically for irreversible reactions. Consequently the constrained null space turned into a bounded convex polyhedron (3).

Artificial Centered Hit and Run (ACHR) method (16-17) was adopted in the sampling progress. For each sampling step in the null space that is constrained by $\alpha$ and $\beta$, given the current point and direction, a line could be drawn through the point along the positive and negative direction. The line was truncated into a segment by the surfaces of the constrained null space, and the next coordinate was generated randomly on this line segment.

To impose the constraint of "loop law", the flux vector which contains unfeasible direction pattern(s) will not be recorded.

*2.3.2. Definition of the Energy Function*

To optimize the Biomass synthesis, the negative value of the Biomass flux was assumed as an energy function:

$$E = -v_{Biomass} \tag{9}$$

Then the optimization turned into the minimization of *E*, and the Metropolis criterion (22) was employed to find the minima of the energy function. When moving every single step in the sampling space, the energy change would be calculated. If the new energy is lower than the old one, the new coordinate would be recorded and the next step would be started from the new coordinate. Otherwise, the acceptance rate *p* would be calculated:

$$p = \exp(\frac{-\Delta E}{T}) \tag{10}$$

Here Δ*E* is the energy change by moving every single step in the sampling, and *T* is a heuristic temperature parameter which controls the variation range of *E* around the minimum. The new coordinate would be accepted by the rate *p*, while by the rate (1 – *p*) it would not be recorded and the next step would start from the old coordinate.

### 2.3.3. Imposition of the Simulated Annealing

For gaining evaluations before choosing acceptance rate and temperature parameter, a random uniform sampling would be performed for certain steps without any rejections, and the energy alternation for each step would be recorded. The initial temperature $T_0$ was chosen as follows (23-24):

$$T_0 = \frac{\Delta E^+}{\ln \frac{n_2}{n_2 p_0 - n_1(1 - p_0)}} \tag{11}$$

where

$$\Delta E^+ = \frac{1}{n_2} \sum_i \frac{\Delta E_i + |\Delta E_i|}{2} \tag{12}$$

Here, $p_0$ is the initial acceptance rate and arbitrarily set to 0.99 to ensure an adequate searching space, $n_1$ and $n_2$ are the numbers of steps in which energy alternations are negative and positive, respectively. Therefore, $\Delta E^+$ represents the average energy change for the steps which cause energy rise.

As an imposition of Simulated Annealing (15), the temperature parameter would be declined to 95% of its previous value after certain steps.

*2.3.4. Optimization of the Biomass Synthesis Efficiency*

However, the previous energy function Eq. 10 disregarded the efficiency of Biomass synthesis, which means that the optimization might be based on a larger amount of the metabolic substrates consumption. To define a criterion to measure the Biomass synthesis efficiency, the Biomass flux was divided by the total amount of reducible electrons in the substrate influxes, and the energy function was modified as:

$$E^u = -\frac{v_{biomass}}{\sum_{i \in R^{red}} v_i \cdot e_i} \qquad (13)$$

Here $R^{red}$ denotes for the set of exchange reactions whose substrates contain reducible electrons, and $e_i$ is the number of reducible electrons of its corresponding substrate. For example, a glucose molecule has 24 reducible electrons compared to its final oxidized products, water and carbon dioxide.

*2.3.5. Consistency of the Sampling Space*

To evaluate the spatial consistency between each two sample tracks, the Cumulative Mean Square Inner Product (CMSIP) (18) was calculated between the eigenvectors of each two sample tracks:

$$CMSIP = \frac{1}{\eta} \sum_{i=1}^{\eta} \sum_{j=1}^{\eta} (\varepsilon_i^A \cdot \varepsilon_j^B)^2 \qquad (14)$$

Here, and $\varepsilon^A$ and $\varepsilon^B$ are the chosen eigenvectors from two different sample trajectories *A* and *B* with largest eigenvalues, and $\eta$ is the number of chosen eigenvectors. The *CMSIP* value varies from 0 to 1, which means that the sampling spaces of the two trajectories vary from no intersection to identical, and the value above a threshold of 0.5 could be considered as significantly overlapped (18).

In some cases, however, the variances represented by the chosen number of eigenvectors are different from the other, which makes the results incomparable. In the circumstance where the eigenvalues of the chosen eigenvectors occupy less proportion of the total eigenvalues, even larger *CMSIP* values tend to be less convincing. In compensation, the final criterion of consistency evaluation in this work was modified by multiplying *CMSIP* with the cumulative proportion of the corresponding eigenvalues.

$$C = CMSIP \cdot P_\eta^A \cdot P_\eta^B \tag{15}$$

Where $P_\eta^A$ means the proportion of the largest $\eta$ eigenvalues in the sum of all the eigenvalues of trajectory A.

**2.4. Minimizing the Free Futile Flux**

As described before, the Futile Pathways have exchange reactions. The fluxes in a certain part of Futile Pathways, whose substrates contain reducible electrons, could be constrained by the energy function Eq. 13, which serves to minimize the energy waste during the Biomass synthesis. However, the fluxes in the other pathways, named Free Futile Pathways, could only be constrained by the biological capacity and result in large variations of flux values. These variations might lead to cover the correlation between the fluxes in functional pathways. To locate the reaction combinations of the FFP, two series of Linear Programming have been performed. The progress is similar to Eq. 8 and Eq. 9, while the difference is that the fluxes in $R^{red}$ were forced to zero, instead of $R^{ex}$. The set of the reaction combinations of the FFP were denoted as $R^f$.

To minimize the futile fluxes (fluxes in these FFP) in a certain sample track, a Linear Programming based algorithm was developed:

$$\min \sum_{i \in R_j^f} |v_i^f|$$

$$s.t. \begin{cases} S \cdot v^f = 0 \\ sign(v_i^f) = sign(v_i), i \in R \\ \alpha_i \leq v_i^f \leq \beta_i, i \in R \\ v_k^f = v_k, k \in \overline{R_j^f} \end{cases} \tag{16}$$

Here $R_j^f$ is the *j*-th reaction combination of FFP, and $v^f$ corresponds to the flux vector after the minimization of futile fluxes in $R_j^f$.

## 3. RESULT AND DISCUSSION

### 3.1. The Result of Preprocessing

Through the Linear Programming Eq. 1, the flux values of 53 reactions were tightly forced to zero, and afterwards they have been deleted along with the dead-end metabolites.

Through the Linear Programming Eq. 2 and Eq. 3, six internal loops have been located in the original network. Comparing to other methods, such as analyzing extreme pathways and searching for type III pathways (11, 25), Linear Programming consumes much less time and computational resource. Particularly, two of the six loops are composed of the reactions [HPROa and HPROx], and [4HGLS, OCBTi, PHCHGS], respectively. However, all of the five reactions are isolated from the other part of the network, and therefore they have been deleted along with the metabolites that only occur in these reactions.

Through solving the inequalities in Eq. 7, one of the remaining four loops, composed of [ASHERL2, METB1r, SHSL1r, SHSL2r], has no feasible direction pattern that satisfies the thermodynamic constraint. Further analysis indicates that the metabolite hcys-L is synthesized via ASHERL2r and RHCCE, but consumed via SHSL2r only, and the metabolite cyst-L appears only in METB1r and SHSL1r. As a result of that, deleting any one of METB1r, SHSL1r, or SHSL2r would cut out other one or more reactions. However, deletion of ASHERL2 would not cause any unfeasible affection of other reactions and might be considered as a better solution to dissemble the loop.

In general, the final version of the network contains 418 metabolites and 499 reactions, including 67 exchange reactions and 3 internal loops:

$R_1^{Loop}$: [ACKr, HSERTA, METB1r, PTAr, SHSL1r, SHSL4r, HSK, THRD_L, THRS];

$R_2^{Loop}$: [H2CO3D, H2CO3D2, HCO3E];

$R_3^{Loop}$: [Nat3_1, PROt2r, PROt4r].

$R_1^{loop}$ has only one feasible direction pattern that satisfies the thermodynamic law, while $R_2^{loop}$ and $R_3^{loop}$ have three and four feasible patterns, respectively.

## 3.2. The Result of Monte - Carlo Sampling

### 3.2.1. Optimization of the Biomass Synthesis Flux

A series of random sampling was performed with imposition of "loop law" and the energy function Eq. 9. Five sample trajectories have been recorded with different random initial points in the sample space, which are chosen by using Linear Programming in a size-shrunk solution space (16). Each trajectory has been recorded into 100,000 samples with an interval of 100 steps, and the temperature parameter was declined every 1,000 samples. The mean values and standard deviations of Biomass synthesis flux of every 10,000 samples are shown in Fig. 2. The results showed well convergence and consistency to each other, and for better confirmation, the modified *CMSIP* (C-value in Eq. 15) values have been calculated between each two adjacent 5,000 samples for every single trajectory with 10 eigenvectors of largest eigenvalues (Fig. 3).

To calculate the mean value of Biomass synthesis flux, the sampling progress was continued by 10,000 samples for each start point and maintained the final temperature unchanged to obtain enough samples under constant tempertures. The C-values among the five 10,000 samples have a mean value of 0.6062 with a standard deviation of 0.0543, which shows a considerable consistency. The mean value of Biomass synthesis flux from the total five 10,000 samples was 1.8587 with a standard deviation of 3.86e-3. In comparison, the Biomass synthesis flux yields at 1.8771 by optimization of Linear Programming. The comparison of mean value and standard deviation of each flux between sampling and Linear Programming was shown in Fig. 4, in which the values of sampling was calculated from the last five 10,000 samples. For a comparable amount of Biomass synthesis, the mean values from the two methods are close to each other; while in sampling producing the same amount of Biomass could bear a certain range of fluctuation for each flux.

### 3.2.2. Optimization of the Biomass Synthetic Efficiency

As a non-linear objective function (Eq. 13), the Biomass synthetic efficiency could not be optimized by Linear Programming. The sampling method here is similar to that in the previous section, but different in declining temperature every 5,000 samples. The results of Biomass synthetic efficiency were shown in Fig. 5, and the C-values were shown in Fig. 6.

Comparing to the result of the optimization of Biomass synthesis flux, the optimization of Biomass synthetic efficiency showed a remarkable increase of synthetic efficiency per each income reducible electron. Similarly each of the five sampling process was continued for 5,000 samples with unchanged temperature, therefore the last 10,000 samples of each trajectory was under constant temperature. The C-values among the five 10,000 samples have a mean value of 0.5840 with a standard variation of 0.0510. The average efficiency of the five 10,000 samples yields 4.172e-3 with a standard deviation of 8.535e-5. In comparison, by calculating the efficiency of previous optimization of Biomass synthesis flux, the mean value is only 8.683e-4 with a standard deviation of 2.276e-5. The efficiency derived from the linear programming is 8.650e-5.

The Fig. 7 represents the comparison of exchange fluxes with reducible electrons between the two optimization methods. The mean values were divided by their corresponding amount of Biomass synthesis for unification. According to the optimization of Biomass synthesis flux, the mean values of exchange exhibit a much wider distribution than those from the optimization of Biomass synthetic efficiency. That means, for synthesizing one unit of Biomass, it consumes more substrate with reducible electrons when optimizing Biomass synthesis flux. A closer comparison between the ingestion and excretion of D-Glucose, Hydrogen, L-Lactate and Succinate is presented in Fig. 8. The results have reaffirmed the increase in substrate usage efficiency.

### 3.3. The Effect of Free Futile Flux Minimization

By using linear programming methods, 15 groups of Free Futile Pathways have been discovered with a total number of 71 reactions. The reactions of EX_co2(e), EX_h2o(e), CO2t and H2Ot participate in all the 15 groups, which serve as tunnels of water and carbon dioxide. The reactions of EX_h(e), EX_nh4(e), EX_urea(e), NH4t, UREAt and UREA participate in 14 groups, which serve as tunnels of ammonia and urea. As a result of that, the 15 groups of FFP could be categorized into two types. The Type I FFP are proton - independent and only contains a carbon acid synthesis pathway (Fig. 9 *A*) in this network, and the Type II FFP are coupled with combinations of proton – driven reactions (Fig. 9 *B*), which serve as proton tunnels.

The last 10,000 samples of the five trajectories of the optimization of Biomass synthetic efficiency were chosen for futile flux minimization (Eq. 16). To demonstrate the effect of minimization, the correlation coefficients between each two reactions before and after the minimization of the 3$^{rd}$ group of FFP were shown in Fig. 10. Each dot refers to a pair of reactions that neither participates in the 3$^{rd}$ group of FFP, each circle refers to a pair that has one reaction in the 3$^{rd}$ group of FFP, and each box refers to a pair of reactions that both participate in the 3$^{rd}$ group of FFP. The dots are aligned diagonally, which proves that the minimization does not affect the correlations among the reactions not in FFP. Some boxes are located at high values of *x*-axis but low values of *y*-axis, which means that the correlations between the reactions in FFP have been weakened. On the contrary, some circles are located at low values of *x*-axis but high values of *y*-axis, which means that the reactions in FFP have gained enhanced correlation with the outside reactions.

Further examination into the 3$^{rd}$ group of FFP could reveal more details. The 3$^{rd}$ group of FFP belongs to Type II and its combination of proton – driven reactions were shown in Fig. 11. Its core reactions, CDAPPA_HP and DASYN_HP, are also involved in the pathway of Glycerophospholipid metabolism (Fig. 12). The correlation coefficients between DASYN_HP and other reactions are shown in Tab. 1, which represents that DASYN_HP was highly coupled with CDAPPA_HP, CYTK1 and NDPK3 before the minimization, and the unreasonable coupling has been dissolved after the minimization. Contemporaneously, the minimization has tightened the relations between DASYN_HP and its upstream / downstream reactions. The effect of the minimization exerting on the mean values and standard variations of CYTK1, DASYN_HP, NDPK3 and CDAPPA_HP are shown in Fig. 13. Especially, the flux values of CDAPPA_HP were almost forced to zero, which indicates an elimination of the coupling loop of CDAPPA_HP and DASYN_HP.

## 4. CONCLUSION

In summary, a series of network simulation and analysis methods have been presented here, based on Monte – Carlo sampling and Linear Programming. The advantages of sampling approach are demonstrated via the imposition of non - linear constraints, like the "loop law", and non – linear objective of optimization. Energy function and Metropolis criterion were introduced to optimize the Biomass synthetic flux and Biomass synthetic efficiency. The temperature factor plays an important role during the optimization, for a higher temperature could lead to wider sample distribution but might result in non – optimization, while a lower temperature could reach a lower zone of energy but might cause the samples trapped into local minima. To solve this dilemma, the Simulated Annealing method has been employed in the sampling, and has been proved effective in the optimizations.

Moreover, to extenuate the dependence on the constraints of artificial biological capacity, the optimization of Biomass synthetic efficiency has been proposed. Comparing to the flux minimization (6), our model has 67 exchange reactions, which are all reversible reactions. Therefore, a number of $2^{67}$ in/out combinations should be taken into account when using linear programming, while the combinations could be automatically chosen and decided during the sampling progress. For the incapability of acquiring the $\Delta G$ of each substrate, the unification criterion has been altered to be the total number of income reducible electrons. The optimization result showed an improved usage efficiency of metabolic substrates and a lower level of by-product excretion than the optimized result of Biomass synthesis flux.

In addition, the concept of Futile Pathway has been proposed and a Linear Programming based method has been developed to minimize the fluxes in Free Futile Pathways. A specific group of FFP has been studied, and the minimization of the fluxes in this group of FFP has shown a significant improvement in two biological aspects. One aspect is that the couplings of reactions in the FFP have been dissolved, and the relations of the reactions which form meaningful biological pathways have been enhanced. The other one is reflected by the decreased flux values of the reactions in FFP, which means that the unnecessary consumed substrates have been economized while the Biomass synthesis was kept at the same value.

**Reference**

1. Amit Varma, and B. O. Palsson. 1994. Metabolic Flux Balancing: Basic Concepts, Scientific and Practical Use. Bio/Technology 12:994-998.
2. Raman, K., and N. Chandra. 2009. Flux balance analysis of biological systems: applications and challenges. Brief Bioinform.
3. Schilling, C. H., S. Schuster, B. O. Palsson, and R. Heinrich. 1999. Metabolic pathway analysis: basic concepts and scientific applications in the post-genomic era. Biotechnol Prog 15:296-303.
4. Edwards, J. S., R. U. Ibarra, and B. O. Palsson. 2001. In silico predictions of Escherichia coli metabolic capabilities are consistent with experimental data. Nature biotechnology 19:125-130.
5. Mahadevan, R., and C. H. Schilling. 2003. The effects of alternate optimal solutions in constraint-based genome-scale metabolic models. Metabolic engineering 5:264-276.
6. Holzhutter, H. G. 2004. The principle of flux minimization and its application to estimate stationary fluxes in metabolic networks. Eur J Biochem 271:2905-2922.
7. Schellenberger, J., and B. O. Palsson. 2009. Use of randomized sampling for analysis of metabolic networks. The Journal of biological chemistry 284:5457-5461.
8. Thiele, I., T. D. Vo, N. D. Price, and B. O. Palsson. 2005. Expanded metabolic reconstruction of Helicobacter pylori (iIT341 GSM/GPR): an in silico genome-scale characterization of single- and double-deletion mutants. J Bacteriol 187:5818-5830.
9. Beard, D. A., S. D. Liang, and H. Qian. 2002. Energy balance for analysis of complex metabolic networks. Biophys J 83:79-86.
10. Beard, D. A., E. Babson, E. Curtis, and H. Qian. 2004. Thermodynamic constraints for biochemical networks. J Theor Biol 228:327-333.
11. Price, N. D., I. Famili, D. A. Beard, and B. O. Palsson. 2002. Extreme pathways and Kirchhoff's second law. Biophys J 83:2879-2882.
12. Price, N. D., I. Thiele, and B. O. Palsson. 2006. Candidate states of Helicobacter pylori's genome-scale metabolic network upon application of "loop law" thermodynamic constraints. Biophys J 90:3919-3928.
13. Brown, K. S., C. C. Hill, G. A. Calero, C. R. Myers, K. H. Lee, J. P. Sethna, and R. A. Cerione. 2004. The statistical mechanics of complex signaling networks: nerve growth factor signaling. Physical biology 1:184-195.
14. Mendes, P., and D. Kell. 1998. Non-linear optimization of biochemical pathways: applications to metabolic engineering and parameter estimation. Bioinformatics 14:869-883.
15. Kirkpatrick, S., C. D. Gelatt, Jr., and M. P. Vecchi. 1983. Optimization by Simulated Annealing. Science 220:671-680.

16. Thiele, I., N. D. Price, T. D. Vo, and B. O. Palsson. 2005. Candidate metabolic network states in human mitochondria. Impact of diabetes, ischemia, and diet. The Journal of biological chemistry 280:11683-11695.
17. Kaufman, D. E., and R. L. Smith. 1994. Direction choice for accelerated convergence in hit-and-run sampling. Operations Research 46:84-95.
18. de Groot, B. L., D. M. van Aalten, A. Amadei, and H. J. Berendsen. 1996. The consistency of large concerted motions in proteins in molecular dynamics simulations. Biophys J 71:1707-1713.
19. Papin, J. A., J. L. Reed, and B. O. Palsson. 2004. Hierarchical thinking in network biology: the unbiased modularization of biochemical networks. Trends Biochem Sci 29:641-647.
20. Amieva, M. R., and E. M. El-Omar. 2008. Host-bacterial interactions in Helicobacter pylori infection. Gastroenterology 134:306-323.
21. Famili, I., and B. O. Palsson. 2003. Systemic metabolic reactions are obtained by singular value decomposition of genome-scale stoichiometric matrices. J Theor Biol 224:87-96.
22. Metropolis, N., A. W. Rosenbluth, M. N. Rosenbluth, and A. H. Teller. 1953. Equation of State Calculations by Fast Computing Machines. Journal of Chemical Physics 21:1087-1092
23. Van Laarhoven, P. J. M., and E. H. L. Aarts. 1987. Simulated Annealing: Theory and applications. D. Reidel.
24. Aarts, E. H. L., and J. H. M. Korst. 1989. Simulated annealing and Boltzmann machines: A stochastic approach to combinatorial optimization. John Wiley & Sons.
25. Bell, S. L., and B. O. Palsson. 2005. Expa: a program for calculating extreme pathways in biochemical reaction networks. Bioinformatics 21:1739-1740.

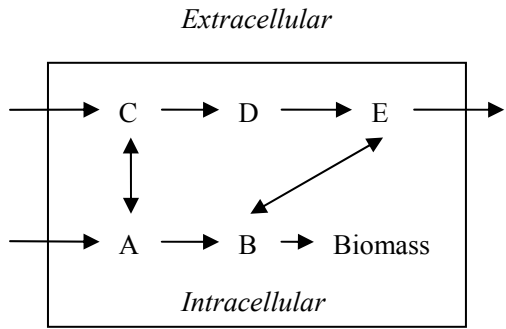

**FIGURE 1** Sketch map of futile pathways

This hypothetic network has four futile pathways:

→ C → D → E →,

→ A → B → E →,

→ C → A → B → E →,

→ A → C → D → E →.

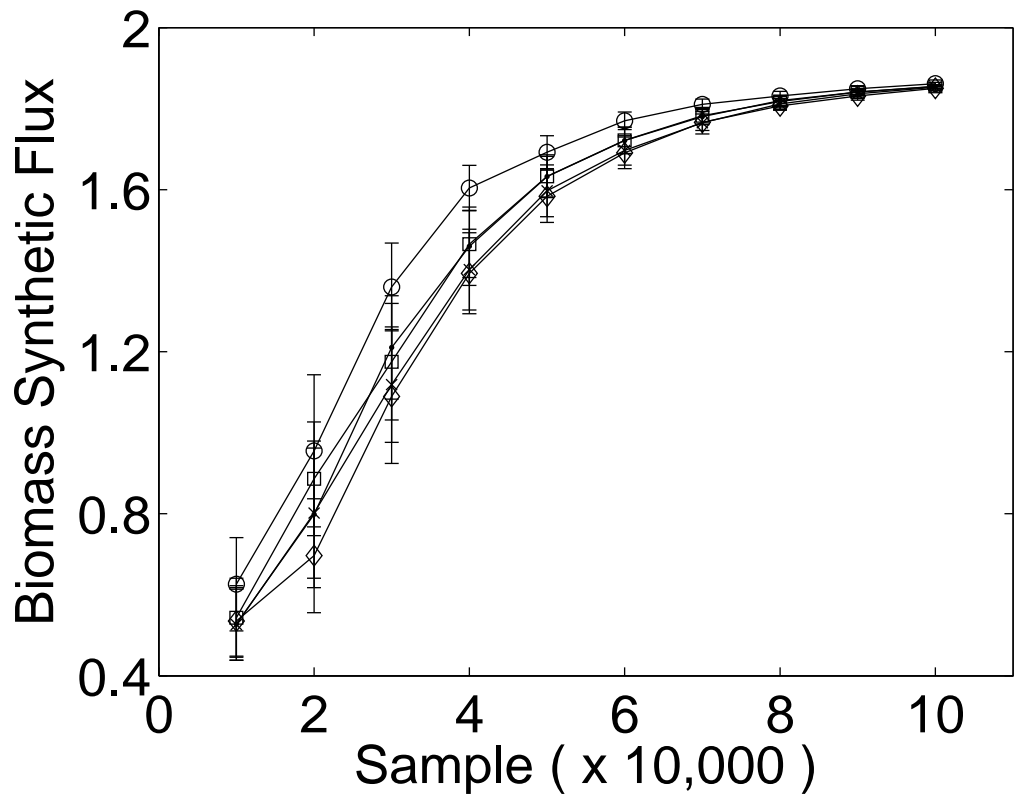

**FIGURE 2** Mean values and standard deviations of the Biomass synthetic flux of the five sample trajectories with different random start points. Each mean value (indicated by circle, rectangular, diamond, asterisk and dot) and standard deviation (indicated by bar) was calculated for the previous 10,000 samples anterior to its corresponding horizontal axis value.

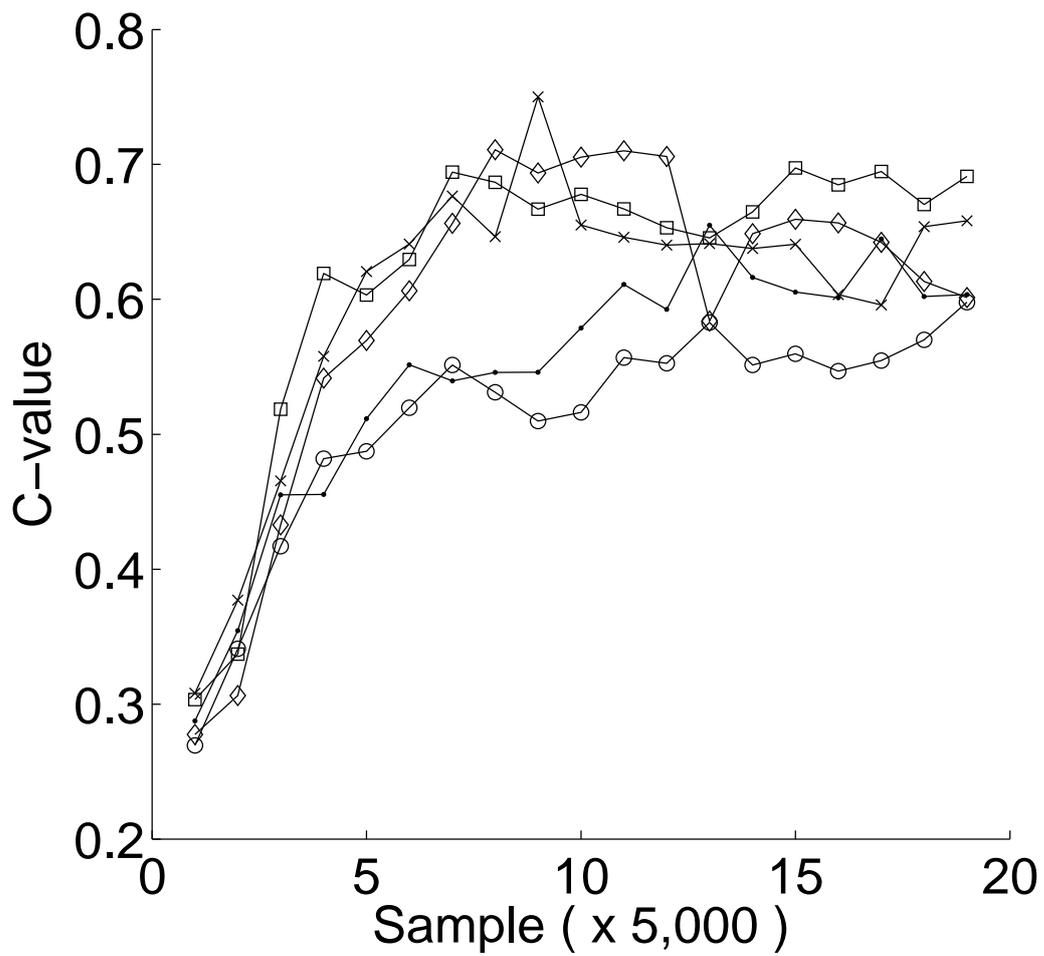

**FIGURE 3** C-value of each trajectory of optimizing the Biomass synthesis flux. The value of each point is calculated by its neighboring two factions of each 5,000 samples with 10 eigenvectors of largest eigenvalues.

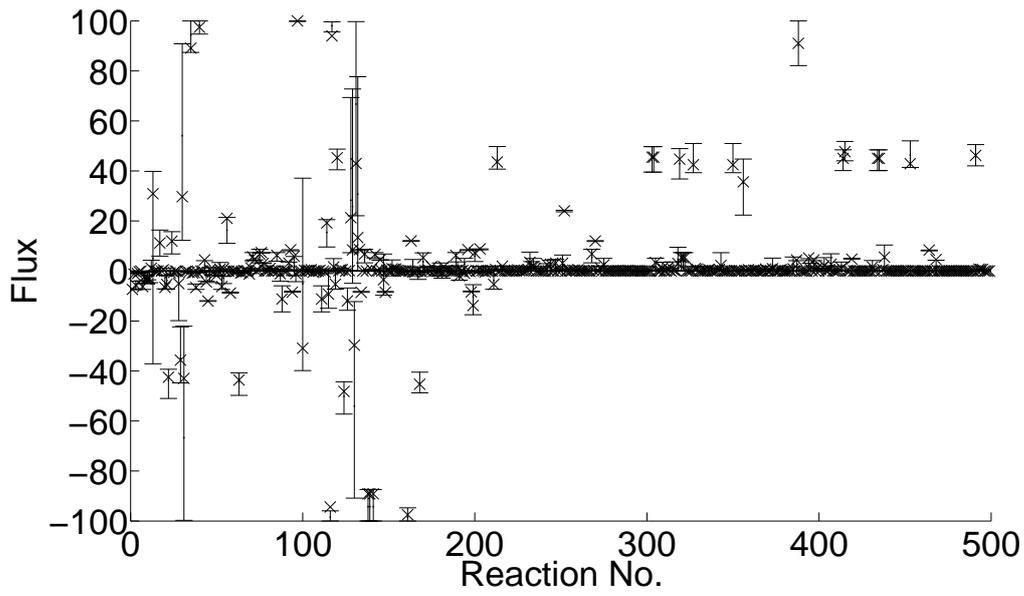

**FIGURE 4** Comparison between the optimization result of sampling and Linear Programming.

Dot and bar: Mean value and standard deviation of each flux from sampling.

Cross: flux values from Linear Programming.

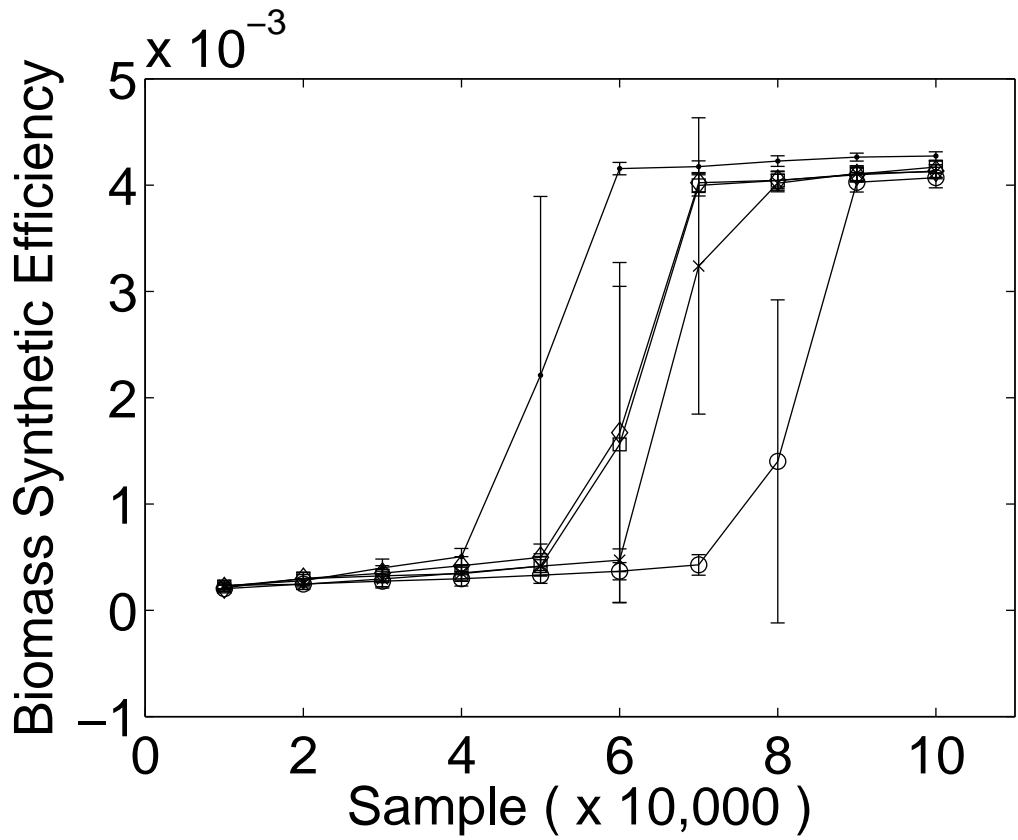

**FIGURE 5** Mean values and standard deviations of the Biomass Synthetic Flux of the five sample trajectories with different random start points. Each mean value (indicated by circle, rectangular, diamond, asterisk and dot) and standard deviation (indicated by bar) was calculated for the previous 10,000 samples anterior to its corresponding horizontal axis value.

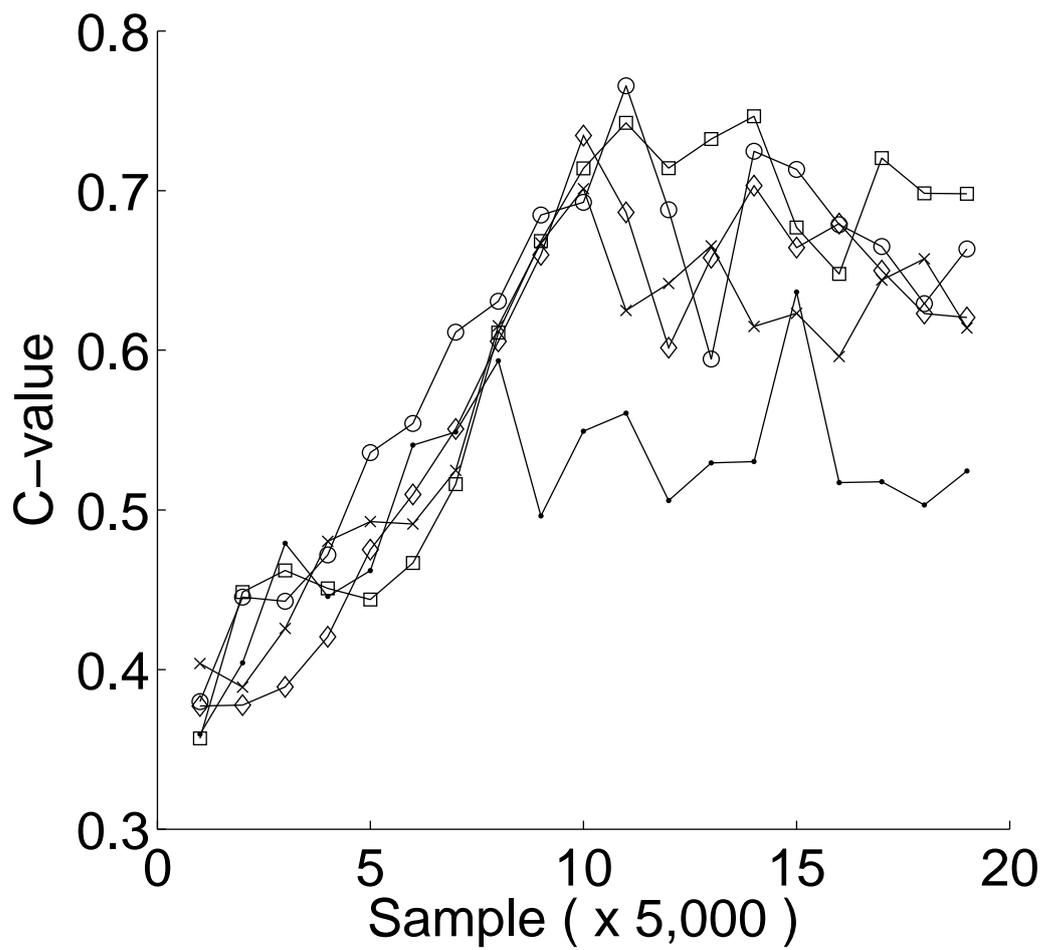

**FIGURE 6** C-value of each trajectory of optimizing the Biomass synthesis efficiency. The value of each point is calculated by its neighboring two factions of each 5,000 samples with 10 eigenvectors of largest eigenvalues.

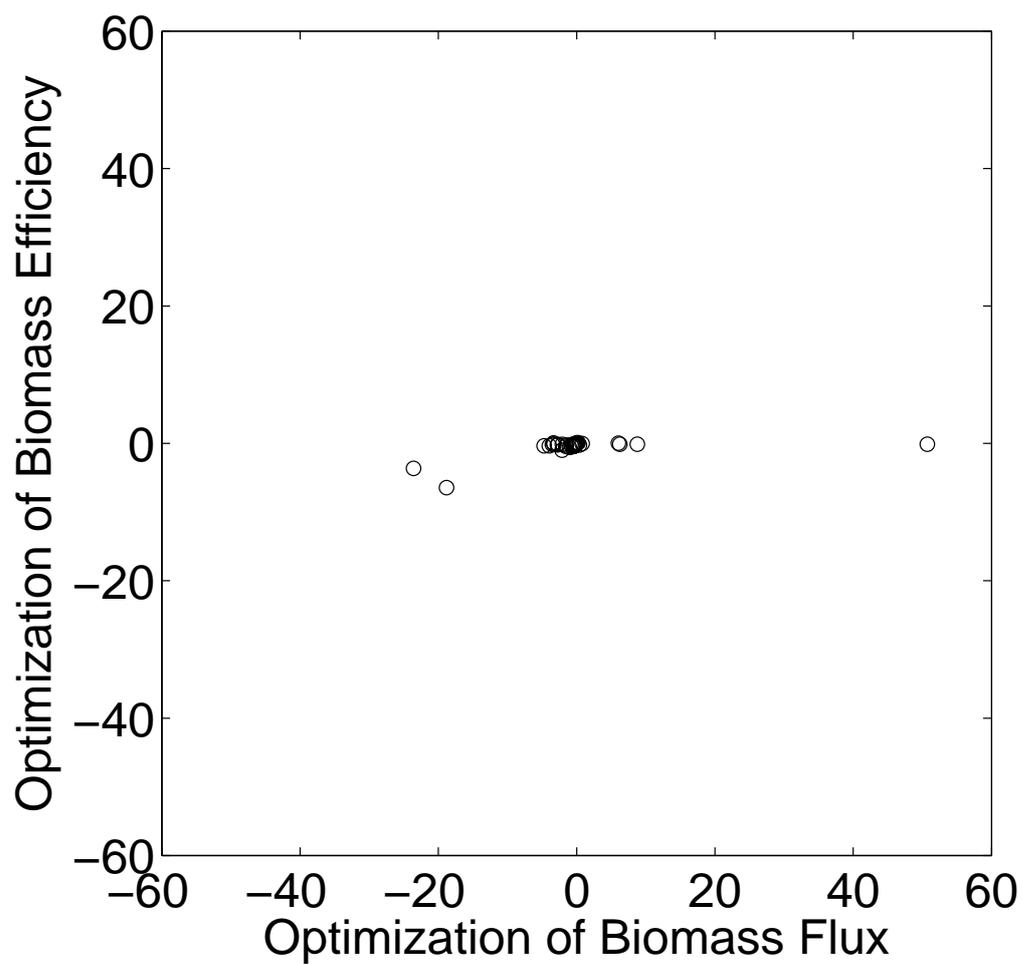

**FIGURE 7** Comparison of the mean value of every exchange flux divided by the Biomass synthesis flux. Each circle represents the unified mean value of an exchange flux, while its corresponding horizontal axis value represents the mean value in optimizing Biomass synthesis flux and its corresponding vertical axis value represents the mean value in optimizing Biomass synthesis efficiency.

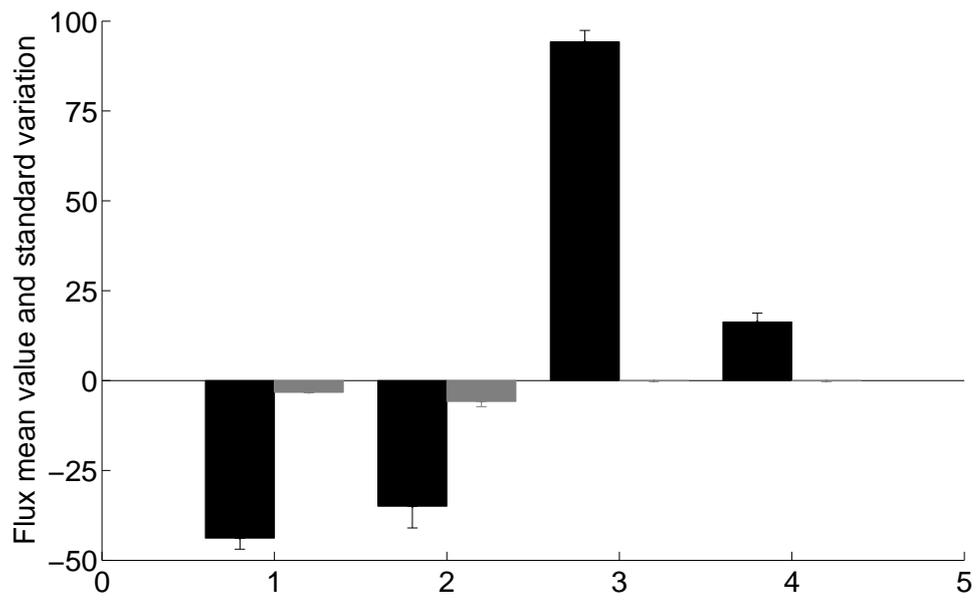

**FIGURE 8** Comparison of four exchange reactions, unified by the amount of Biomass. Positive value means excretion and negative value means ingestion.
Black: optimizing Biomass synthetic flux; Grey: optimizing Biomass synthetic efficiency.
Column 1: D-Glucose; Column 2: Hydrogen; Column 3: L-Lactate; Column 4: Succinate.

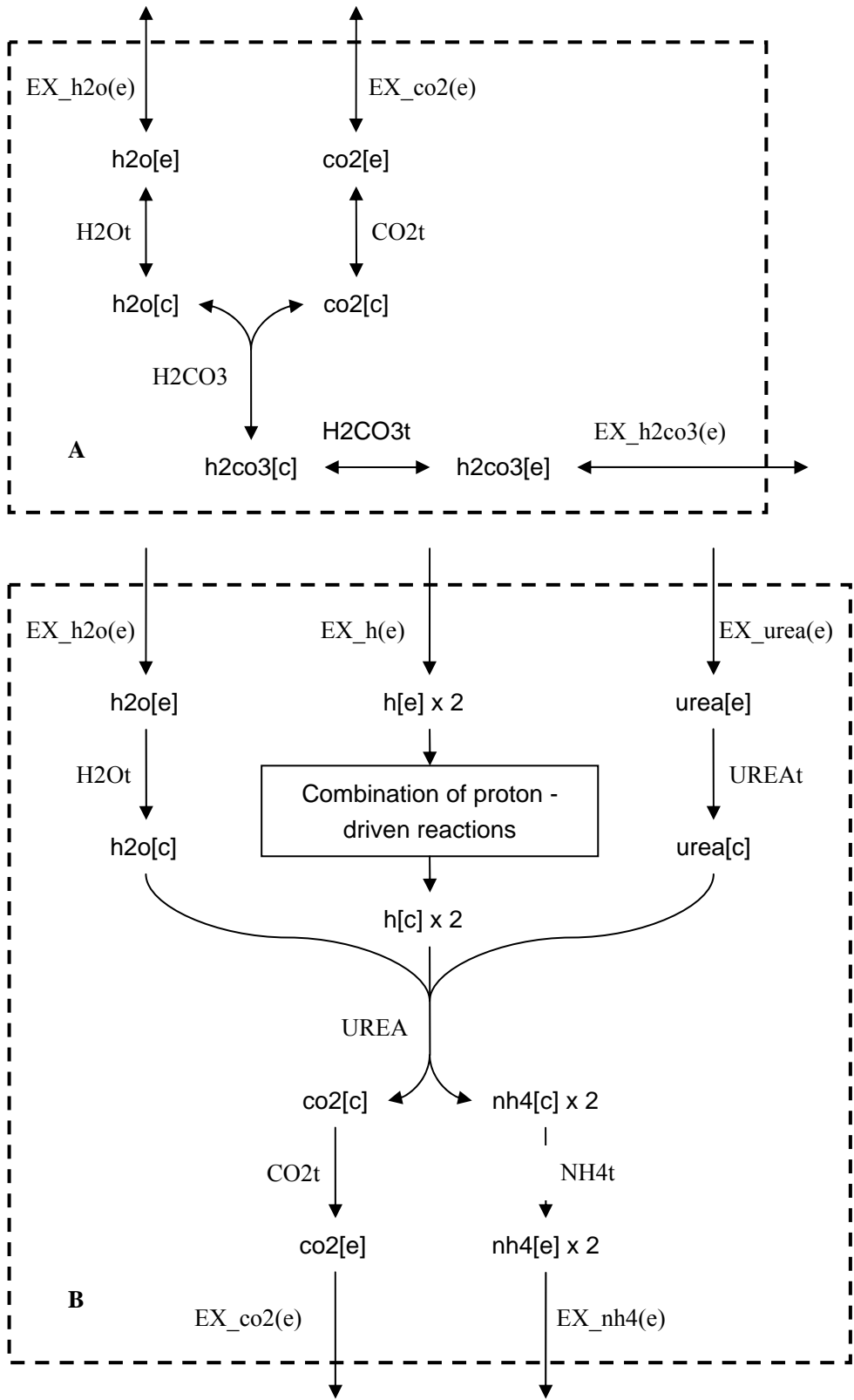

**FIGURE 9** Two types of Free Futile Pathways
A. carbonic acid synthesis (proton - independent)
B. proton - dependent pathways

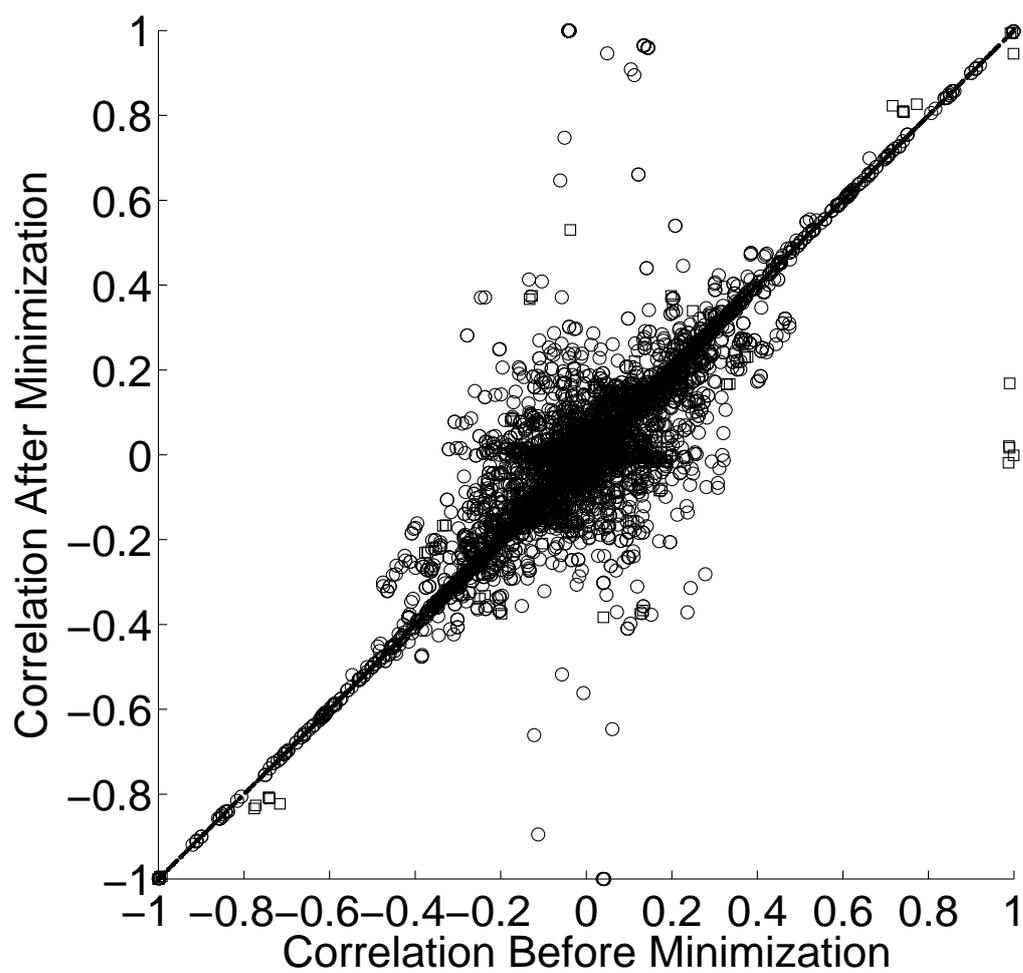

**FIGURE 10** Comparison of correlation coefficients between each two reactions, before and after the minimization of fluxes in the 3$^{rd}$ FFP.

Dots: both two reactions do not participate in the 3$^{rd}$ group of FFP.
Circles: one of the two reactions participates in the 3$^{rd}$ group of FFP.
Boxes: both two reactions participate in the 3$^{rd}$ group of FFP.

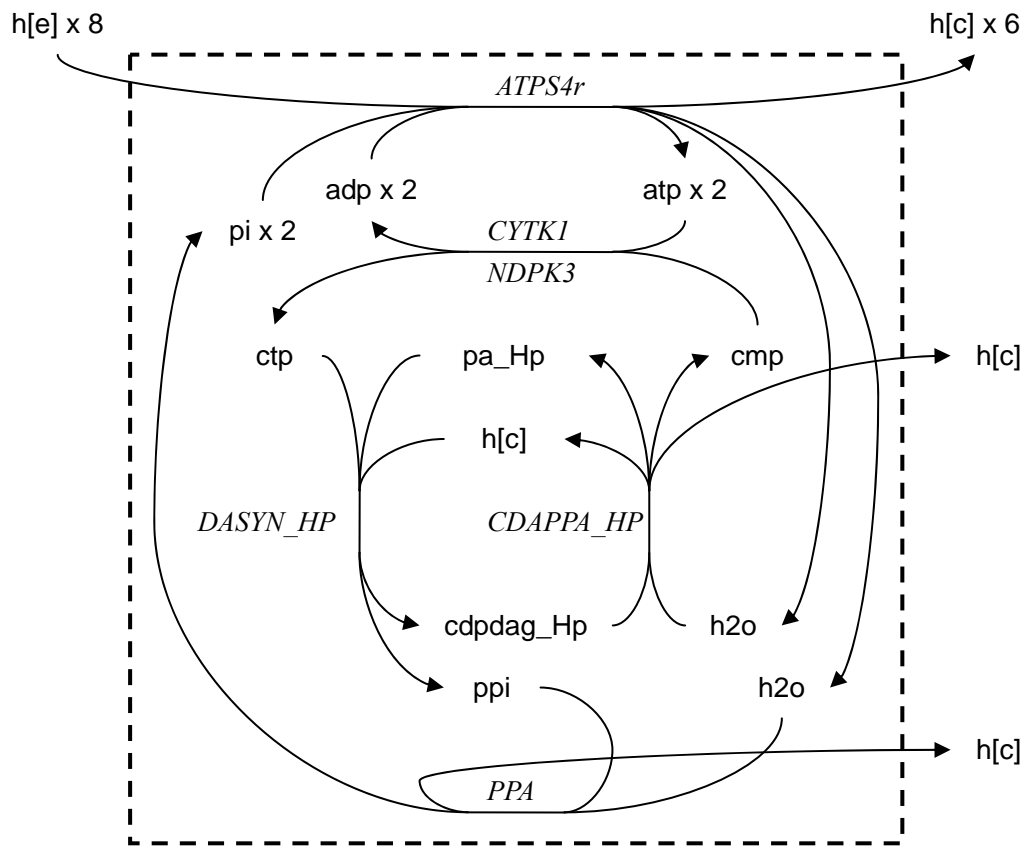

**FIGURE 11** The combination of proton – driven reactions in the 3${}^{rd}$ group of FFP.

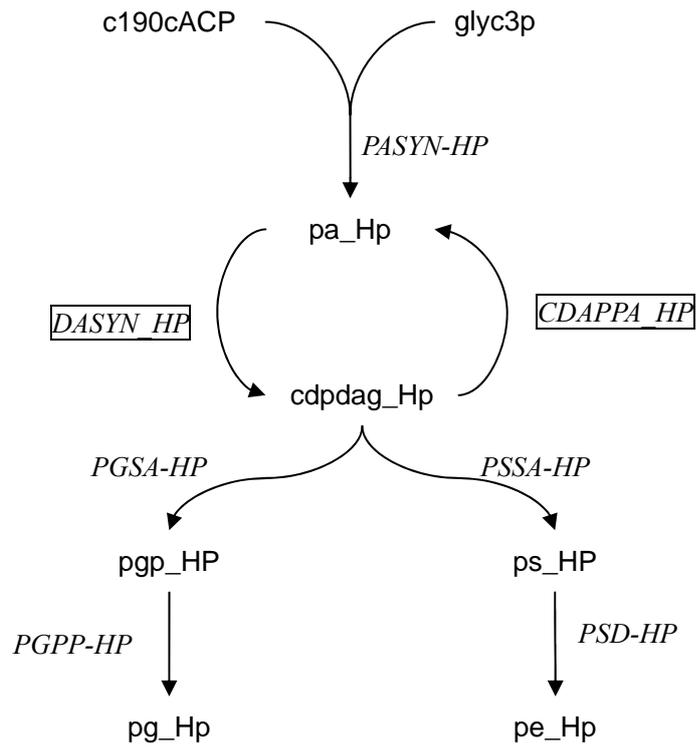

**FIGURE 12** Part of the pathway of Glycerophospholipid metabolism in *H. pylori*. The two boxed reactions participate in the 3$^{rd}$ group of FFP.

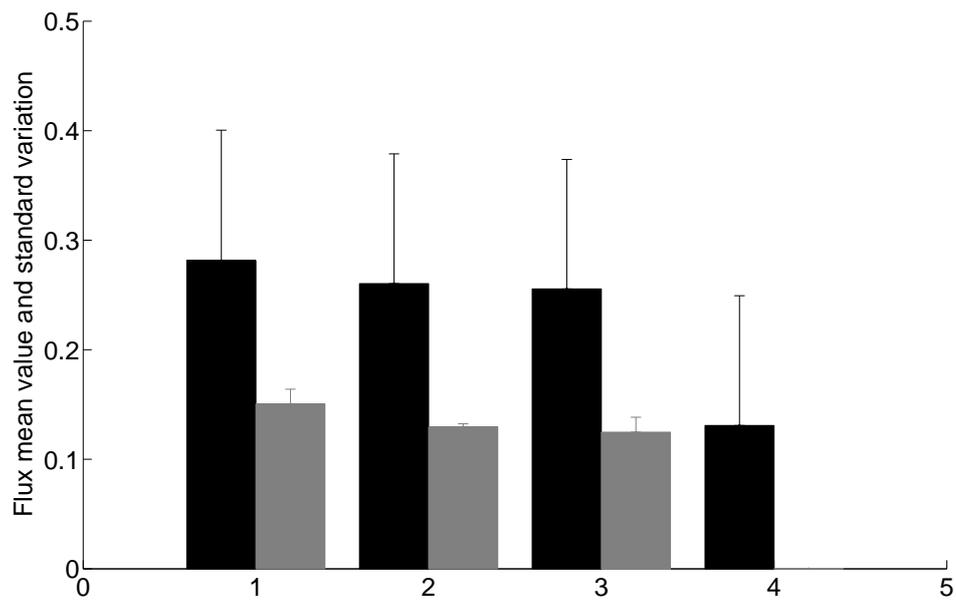

**FIGURE 13** The comparison of mean values and standard variations of four reactions in the $3^{rd}$ group of FFP.

Black: before the minimization; Grey: after the minimization.

Column 1: CYTK1; Column 2: DASYN_HP; Column 3: NDPK3; Column 4: CDAPPA_HP.

**Table 1. The correlation coefficients before and after imposing EBF**

|  | DASYN_HP Before minimization | DASYN_HP After minimization | Memo |
|---|---|---|---|
| CYTK1 | 0.9900 | 0.1682 | Both in futile pathway |
| NDPK3 | 0.9874 | -0.0186 | |
| CDAPPA_HP | 0.9994 | -0.0011 | |
| PASYN_HP | -0.0415 | 1.0000 | Only DASYN_HP in futile pathway |
| PGSA_HP | -0.0415 | 1.0000 | |
| PSSA_HP | -0.0415 | 1.0000 | |
| PGPP_HP | -0.0415 | 1.0000 | |
| PSD_HP | -0.0415 | 1.0000 | |
| Biomass | -0.0415 | 1.0000 | |